\documentstyle[12pt]{article}

\newcommand{\be}{\begin{equation}}
\newcommand{\ee}{\end{equation}}
\newcommand{\eee}{\end{eqnarray}}
\newcommand{\bee}{\begin{eqnarray}}

\newcommand{\gal}{\alpha}
\newcommand{\gb}{\beta}
\newcommand{\gga}{\gamma}

\newcommand{\gl}{\lambda}

\newcommand{\gep}{\epsilon}
\newcommand{\gvep}{\varepsilon}
\newcommand{\gs}{\sigma}
\newcommand{\go}{\omega}

\newcommand{\ty}{\hat{y}}
\newcommand{\nn}{\nonumber}
\newcommand{\half}{\frac{1}{2}}

\begin{document}

\thispagestyle{empty}

\begin{flushright}
\vspace{1mm}
FIAN/TD/27--98\\
{December 1998}\\
\end{flushright}

\vspace{1cm}

\centerline{\large\bf{{3D HIGHER-SPIN GAUGE THEORIES WITH MATTER}}${\
}^{\dag}$}

\begin{center}
{}
\vspace{2cm}
{\bf Sergey Prokushkin and Mikhail Vasiliev }
\vspace{1cm}

I.E.Tamm Department of Theoretical Physics, Lebedev Physical
Institute,\\
Leninsky prospect 53, 117924, Moscow, Russia
\vspace{1.5cm}
\end{center}

\begin{abstract}
This paper is a letter-type version of hep-th/9806236.
We discuss properties of non-linear equations of motion which
describe higher-spin gauge interactions for massive spin-0 and
spin-1/2 matter fields in 2+1 dimensional anti-de Sitter space.
The model is shown to have $N=2$ supersymmetry and to describe
higher-spin interactions of  $d3$ $N=2$ massive hypermultiplets.
An integrating flow is found which reduces the full non-linear
system to the free field equations via a non-local
B\"acklund-Nicolai--type mapping.

\end{abstract}

\vskip5cm

${\ }^{\dag}$
Based on the talks given at the
International Seminar "Supersymmetries and Quantum Symmetries"
in memory of Professor  V.~I.~Ogievetsky (June 22 - 26, 1997,
Dubna, Russia) and the 31st International Symposium Ahrenshoop
(September 2 - 6, 1997, Buckow, Germany).

\newpage

\section{Introduction and Preliminaries}

In this paper, we summarize some recent results in the study
of the higher-spin (HS) interactions of massive matter fields
in 2+1 dimensional space-time.
A most important argument to consider
 theories of HS gauge fields is that they may open
an alternative way towards a fundamental theory presently identified
with $M$-theory. Although a route to $M$-theory  passes through
models in higher dimensions, particularly $d=11$ and $d=12$,
it is useful to study a $d=3$ model
which has much simpler dynamics
because $d3$ HS gauge fields do not propagate \cite{bl}.
As shown below, the simplicity of the model indeed allows one to
study it in great detail,  leading to some general
conclusions on a structure of more complicated HS models.
The main  result consists of the  constructive definition
of a non-linear non-local mapping
%\cite{BN}
which links  the non-linear problem with the free one.
Due to the analogy of the $d=3$ models with
the self-dual $d=4$ models \cite{more} we speculate that the
existence of the integrating flow may be an indication of
integrability of the model.

It is convenient to describe HS gauge fields within
``geometric approach" to gravity
\cite{U,Kib,ChW,MM,SW} with vielbein $h_{\mu}{}^{a}$ and
Lorentz connection $\omega_\mu{}^{ab}$ identified with the
connection 1-forms of an appropriate space-time symmetry algebra $g$.
For example, one can use gauge fields
$A_\mu^{BC}=-A_\mu^{CB}$ of the anti-de Sitter (AdS)
algebra $g = o(d-1,2)$ to describe the geometry of the
$d$-dimensional AdS space-time
(the indices $B,C=0,...,d$ are raised and lowered by the flat
metrics $\eta^{BC} = diag(+- \cdots -+)$)
setting $\go_{\mu}{}^{ab}=A_\mu^{ab}$ and
$h_{\mu}{}^a =(\sqrt{2}\gl)^{-1} A_\mu^{a\, \cdot }$
with the conventions $a,b = 0,...,d-1$, and $B=(b,\,\cdot)$.
Here $\gl \neq 0$ is some constant.
The respective $o(d-1,2)$ gauge curvatures have the form
\be
\label{rim1}
   R_{\mu\nu}{}^{ab}=\partial_{\mu}\go_{\nu}{}^{ab}
     +\go_{\mu}{}^a{}_c\,\go_{\nu}{}^{cb}
     -2\gl^2 h_{\mu}{}^a\,h_{\nu}{}^b - (\mu\leftrightarrow \nu) \,,
\ee
\be
\label{tor1}
   R_{\mu\nu}{}^a=\partial_{\mu}h_{\nu}{}^a
     +\go_{\mu}{}^a{}_c h_{\nu}{}^c - (\mu\leftrightarrow \nu) \,.
\ee
Lorentz connection $\go_{\mu}{}^{ab}$ is expressed
via vielbein $h_{\mu}{}^a$ with the aid of the constraint
$R_{\mu\nu}{}^a=0$ ($h_{\mu}{}^a$ is assumed to be non-degenerate).
Substituting $\go_{\mu}{}^{ab} = \go_{\mu}{}^{ab}(h)$ into
(\ref{rim1}),
 the equation $R_{\mu\nu}{}^{ab}=0$ becomes equivalent to
${\cal R}_{\mu\nu}{}^{ab}
=2\gl^2(h_{\mu}{}^a\,h_{\nu}{}^b-h_{\nu}{}^a\,h_{\mu}{}^b) \,,$
where ${\cal R}_{\mu\nu}{}^{ab}=\partial_{\mu}\go_{\nu}{}^{ab}(h)
+\go_{\mu}{}^a{}_c(h)\,\go_{\nu}{}^{cb}(h)-(\mu\leftrightarrow \nu)$
is the Riemann tensor.
Therefore, the equations $R_{\mu\nu}{}^{ab}=0$ and $R_{\mu\nu}{}^a=0$
describe AdS space-time with radius $(\sqrt{2}\gl)^{-1}$.
This is how AdS space appears as a vacuum
solution of the HS equations considered below.
A role of the algebra
$o(d-1,2)$ is twofold: its
connection 1-forms are identified with the dynamical
fields of the theory and it
serves as the symmetry algebra of the most symmetric vacuum solution,
AdS space-time symmetry.

In the case of $d=2+1$, the AdS algebra is
$o(2,2)$, and the gravitational action
is the Chern-Simons action,
$ S^W =\int_{M_3} str (w\wedge dw +\frac{2}{3}w\wedge w\wedge w) $,
where $w$ is the $o(2,2)$ connection 2-form \cite{W}.

The $d=2+1$ HS superalgebras used below can be
described as follows. Let $Aq(2;\nu)$ be an associative algebra
\cite{Op} with a general element of the form
\be
\label{gel}
  f(\ty, k)\!\! =  \!\!\!\! \sum^\infty_
    {\tiny {\begin{array}{c} n=0\\A=0,1 \end{array}}} \!\!\!
    \frac{1}{n!}  f^{A\,\gal_1\ldots\gal_n}(k)^A
    \ty_{\gal_1}\ldots \ty_{\gal_n}\,,
\ee
under condition that the coefficients
$ f^{A\,\gal_1\ldots\gal_n}$ are symmetric with respect to
the indices $\gal_j =1,2$, while the generating elements
$\ty_\gal$, $k$ satisfy the relations
\bee
\label{modosc}
   [\ty_\gal, \ty_\gb ] = 2i\gep_{\gal\gb} (1+\nu k)\,,
   \quad k \ty_\gal = -\ty_\gal k\,, \quad k^2 = 1\,,
\eee
where $\nu$ is an arbitrary number
($\gal , \,\gb\,,\gga\, =1,2$ are
$d=2+1$ spinor indices lowered and raised by
 the symplectic form $\gep_{\gal\gb}=-\gep_{\gb\gal}$,
$\gep_{ 12} =\gep^{12}=1$,
$A^\gal =\gep^{\gal\gb}A_\gb$,
$A_\gal = A^\gb \gep_{\gb\gal}$).
Then, the $d3$ HS superalgebra $hs(2;\nu)$ is the Lie superalgebra
canonically related to the associative algebra $Aq(2;\nu)$
(i.e. a product law in  $hs(2;\nu)$ is identified with
(anti)commutators in
$Aq(2;\nu)$ for the $Z_2$ grading counting a number of
spinor indices,
i.e. $\pi (\hat{y}_\gga ) =1$, $\pi (k) =0$).
To describe a doubling of the elementary algebras in
$g = hs(2;\nu )\oplus hs(2;\nu )$ analogous to
$o(2,2)\sim sp(2)\oplus sp(2)$
it is convenient to introduce an additional central
involutive generating element $\psi$,
\be
   [\psi , \ty_\gal ]=0\,,\quad [\psi, k]=0\,,\quad \psi^2 =1\,.
\ee
The two simple components of $g$ are singled out by
the projection operators $P_\pm$ = $\half (1\pm \psi )$.
Field strengths for the gauge algebra $g$ are
\be
   R(\ty,\psi,k|x) = d\go(\ty,\psi,k|x) -
   \go(\ty,\psi,k|x) \wedge \go(\ty,\psi,k|x) \,,
\ee
where $d=dx^\nu \frac{\partial}{\partial x^\nu} $ and
$\go(\ty,\psi,k|x)$ are the gauge fields for $g$ of the form
(\ref{gel}) depending also on $\psi$ and the commuting space-time
coordinates $x^\mu$, $\mu=0,1,2$.

The $d3$ AdS space-time symmetry algebra
$o(2,2)\sim sp(2)\oplus sp(2)$ is the subalgebra of $g$
spanned by the bilinears
\be
\label{al}
   L_{\gal\gb}=\frac1{4i}\{\hat{y}_\gal,
    \hat{y}_\gb\}\,,\qquad \,
   P_{\gal\gb}=\frac1{4i}\{\hat{y}_\gal,
    \hat{y}_\gb\}\psi  \,.
\ee

The pure gauge HS action has the Chern-Simons
form with the supertrace defined in \cite{Op}.
It reduces to the Witten gravity action
\cite{W} in the spin 2 sector and to the Blencowe's HS action
\cite{bl} in the case  $\nu =0$.

An important question we address below at the level of
equations of motion is how to introduce
interactions of HS gauge fields with propagating matter fields.
We will follow the ``unfolded formulation"  of
\cite{un}, rewriting dynamical equations in a form of certain
zero-curvature conditions and covariant constancy conditions
\be
\label{0cur}
   d\go =\go\wedge\go\,,\qquad
   dB^A =\go^i (t_i){}^A{}_B B^B\,,
\ee
supplemented with some gauge invariant constraints
\be
\label{con}
   \chi (B) =0
\ee
that do not contain space-time derivatives.
Here  $\go (x)=dx^\nu \go_\nu^i (x) T_i $ is a gauge field
of some Lie superalgebra $l$ ($T_i \in l)$,
and $B^A (x)$ is a set of 0-forms that take values in
a representation space of
some representation $(t_i ){}^B {}_A$ of $l$.

An interesting property of this form of the
equations is that their dynamical content
is hidden in the constraints (\ref{con}).
Indeed, locally one can integrate
(\ref{0cur}) as $\go=d g (x)g^{-1} (x)$,
$B(x)=t_{g(x)} (B_0)$, where $g(x)$ is an arbitrary invertible
element, while $B_0$ is an arbitrary $x$ - independent
representation element and $t_{g(x)}$ is the exponential of
the representation $t$ of $l$. Since the constraints $\chi(B)$
are gauge invariant one is left with the only condition
$\label{con0} \chi (B_0)=0$.
Let $g(x_0)=I$ for some point of space-time  $x_0$.
Then $B_0$=$B(x_0)$.
To understand how restrictions on
values of some 0-forms at a fixed point of space-time can lead
to non-trivial dynamics one should take into account that in
the interesting examples %(see \cite{Prop,rev})
the set of 0-forms $B$ is reach enough
to describe all space-time derivatives of dynamical fields,
while the constraints (\ref{con}) just impose all
restrictions on the space-time derivatives required by the
dynamical equations under consideration. Given solution
of (\ref{con}) one knows all derivatives of the dynamical fields
compatible with the field equations and can therefore reconstruct
these fields by analyticity in some neighborhood of $x_0$.

To illustrate this point let us consider an example
of a scalar field $\phi$ obeying the massless Klein-Gordon
equation $\Box\phi=0$ in a flat  space-time
of an arbitrary dimension $d$. Here $l$ is identified with
the Poincar\'e algebra $iso(d-1,1)$ which gives rise to
the gauge fields
$\go_\nu =(h_\nu{}^a , \go_\nu {}^{ab})\quad$ ($a,b =0,...,d-1$).
The zero curvature conditions of $iso(d-1,1)$
\be
\label{zerocur}
  R_{\nu\mu}{}^a =0\,,\qquad R_{\nu\mu}{}^{ab}=0
\ee
imply that the vielbein $h_\nu {}^a$
and Lorentz connection $\go_\nu {}^{ab}$ describe the flat geometry.
Fixing the local Poincar\'e gauge transformations one can set
\be
\label{flgauge}
   h_\nu {}^a =\delta _\nu^a \,,\quad \go_\nu{}^{ab}=0\,.
\ee

To describe dynamics of a spin zero massless field
$\phi (x)$ let us introduce an
infinite collection of 0-forms $\phi_{a_1\ldots a_n}(x)$
which are totally symmetric traceless tensors
\be
\label{tr}
   \eta^{bc}\phi_{bca_3\ldots a_n}=0\,,
\ee
where $\eta^{bc}$ is the flat Minkowski metrics.
The ``unfolded" version of the Klein-Gordon equation
has a form of the following infinite chain of equations
\be
\label{un0}
\partial_\nu \phi_{a_1\ldots a_n }(x) =h_\nu {}^b
\phi_{a_1 \ldots a_n b}(x)\,,
\ee
where we have replaced the Lorentz covariant
derivative by the ordinary flat derivative $\partial_\nu$ using
the gauge condition (\ref{flgauge}).
The tracelessness condition (\ref{tr}) is a specific realization of
the constraints (\ref{con}), while the system of equations
(\ref{zerocur}), (\ref{un0}) is a particular example of
the equations (\ref{0cur}).
It is easy to see that this system is formally
consistent, i.e. $\partial_\mu$ differentiation of (\ref{un0})
does not lead to new conditions after antisymmetrization
$\nu\leftrightarrow\mu$. This property is equivalent to the fact that
the set of zero forms $\phi_{a_1 \ldots a_n}$ spans some
representation of the Poincar\'e algebra.

To show that the system (\ref{un0}) is equivalent to the
free massless field equation $\Box \phi (x)=0$ let us identify the
scalar field $\phi (x)$ with the $n=0$ member of the tower
of 0-forms $\phi_{a_1 \ldots a_n}$. Then the first two equations
(\ref{un0}) read
\be
\label{1}
   \partial_\nu \phi =\phi_\nu
\ee
and
\be
\label{2}
   \partial_\nu \phi_\mu= \phi_{\mu\nu}\,,
\ee respectively.
Eq.~(\ref{1}) tells us that
$\phi_\nu$ is a first derivative of $\phi$.
Eq.~({\ref{2}) implies that $\phi_{\nu\mu}$ is a second derivative
of $\phi$. However, because of the tracelessness condition (\ref{tr})
it imposes the Klein-Gordon equation $\Box\phi =0$.
It is easy to see that all other equations in (\ref{un0}) express
highest tensors in terms of the higher-order derivatives
\be
\label{hder}
\phi_{\nu_1 \ldots \nu_n}= \partial_{\nu_1}\ldots\partial_{\nu_n}\phi
\ee
and impose no additional conditions on $\phi$.
The tracelessness conditions
are all satisfied once the Klein-Gordon equation is true.

Let us note that the system (\ref{un0}) without the constraints
(\ref{tr}) remains formally consistent but is dynamically empty just
expressing all highest tensors in terms of derivatives of $\phi$
according to (\ref{hder}). This simple example
illustrates how constraints can be equivalent to the dynamical
equations. The specificity of the HS dynamics considered below
that makes such an approach adequate is that HS
symmetries mix all orders of derivatives which therefore
are contained in a representation space of HS symmetries.

\section{Nonlinear System}

Now let us turn to the system for the massive matter fields
interacting via HS gauge potentials in $d=2+1$.
The full nonlinear system of equations,
which is a particular realization of the equations (\ref{0cur})
and (\ref{con}), is formulated in terms of
the generating functions
$W(z,y;\psi_{1,2},k,\rho | x)$, $B(z,y;\psi_{1,2},k,\rho | x)$,
and $S_\gal(z,y;\psi_{1,2},k,\rho | x)$
that depend on the space-time coordinates $x^\nu$
$(\nu=0,1,2)$, auxiliary commuting spinors $z_\gal$, $y_\gal$
$(\gal=1,2)$,
$[y_\gal,y_\beta]=[z_\gal,z_\beta]=[z_\gal,y_\beta]=0$,
a pair of Clifford elements
$\{\psi_i,\psi_j\}=2\delta_{ij}$ $(i=1,2)$ that commute to
all other generating elements, and another pair of Clifford-type
elements $k$ and $\rho$ which have the following properties
\be
\label{Klein}
   k^2=1\,,\: \rho^2 =1 \,,\: k\rho +\rho k =0\,,\:
   ky_\gal=-y_\gal k\,,\: kz_\gal=-z_\gal k\,, \:
   \rho y_\gal=y_\gal \rho \,,\: \rho z_\gal=z_\gal \rho\,.
\ee

The space-time 1-form $W=dx^\nu W_\nu(z,y;\psi_{1,2},k,\rho | x)$,
\bee
\label{gexp}
   W_\mu(z,y;\psi_{1,2},k,\rho |x) &=& \sum_{A,B,C,D=0}^1
    \sum_{m,n=0}^\infty \!\frac 1{m!n!}W^{ABCD}_{\mu,\,
    {\gal_1}\ldots {\gal_m}{\beta_1}\ldots {\beta_n}}(x) \nn \\
    && {}\times
    k^A \rho^B \psi_1^C \psi_2^D  z^{\gal_1}\ldots z^{\gal_m}
    y^{\beta_1}\ldots y^{\beta_n}
\eee
is the  generating function for HS gauge fields.
$B=B(z,y;\psi_{1,2},k,\rho | x)$
is the generating function for  the matter fields.
The components of its expansion analogous to (\ref{gexp})
are identified with the $d3$ matter fields and all their
on-mass-shell non-trivial derivatives.
$S_\gal(z,y;\psi_{1,2},k,\rho | x)$ describes
auxiliary and pure gauge degrees of freedom.
The multispinorial coefficients
in the expansions like (\ref{gexp}) of the functions
$W_\mu$, $B$, and $S_\gga$ carry standard Grassmann parity
in accordance with the number of spinor indices.

The generating functions are treated as
elements of an associative algebra
with the product law
\be
\label{prod}
   (f*g)(z,y;\psi_{1,2},k,\rho)\!\!
   =\!\!\frac{1}{(2\pi)^2}\!\!\int \!\! d^2ud^2v \!
   \exp(iu_\gal v^\gal)
   f(z+u,y+u;\psi_{1,2},k,\rho)g(z-v,y+v;\psi_{1,2},k,\rho) ,
\ee
where the integration
variables $u$ and $v$ are required to satisfy the commutation
relations similar to those of $y$ and $z$ in (\ref{Klein}).
This product law yields a particular
realization of Heisenberg-Weyl algebra,
$[y_\gal,y_\beta]_*=-[z_\gal,z_\beta]_*=2i\epsilon_{\gal\beta}$,
   $[y_\gal,z_\beta]_*=0$ ($[a,b]_*=a*b-b*a$).

The full system of equations is analogous to
the $d3$ massless system of \cite{Eq},
\be
\label{WW}
      dW=W*\wedge W
\,,\quad
      dB=W*B-B*W
\,,\quad
      dS_\gal=W*S_\gal-S_\gal*W \,,
\ee
\be
\label{SS}
    S_\gal * S_\beta-S_\beta * S_\gal=-2i\epsilon_{\gal\beta}(1+B*K)
\,,\quad
    S_\gal*B=B*S_\gal \,.
\ee
Here $d=dx^\nu \frac{\partial}{\partial x^\nu} $
and $K=k e^{i(zy)}\,,$ $ (zy)=z_\gal y^\gal$.

With the aid of the involutive automorphism $\rho\to -\rho$,
$S_\gal\rightarrow -S_\gal$
one can truncate the system (\ref{WW}), (\ref{SS})
to the one with the
fields $W$ and $B$ independent of $\rho$ and
$S_\gal$ linear in $\rho$,
\be
\label{tru}
    W(z,y;\psi_{1,2},k,\rho | x)=W(z,y;\psi_{1,2},k | x)\,,\quad
    B(z,y;\psi_{1,2},k,\rho | x)=
    B(z,y;\psi_{1,2},k | x)\,,\quad
\ee
\be
    S_\gal(z,y;\psi_{1,2},k,\rho | x)=
      \rho s_\gal(z,y;\psi_{1,2},k | x)\,.
\ee
{}From now on we consider this reduced system.
Eqs. (\ref{WW}), (\ref{SS})
are invariant under the infinitesimal HS gauge transformations
\be
\label{delta W}
       \delta W=d\gvep-W*\gvep+\gvep * W
\,,\quad
       \delta B=\gvep *B-B*\gvep
\,,\quad
       \delta S_\gal=\gvep *S_\gal-S_\gal*\gvep \,,
\ee
where $\gvep=\gvep(z,y;\psi_{1,2},k | x)$ is an arbitrary gauge
parameter.

To elucidate the dynamical content of the system
(\ref{WW}), (\ref{SS}), one first of all has to
find an appropriate vacuum solution.
There exists a class of vacuum solutions \cite{PV}.
The simplest one is
\be
\label{B_0}
   B_0=\nu=const \,,
\ee
\be
\label{S_0}
   S_{0\gal}=\rho \left(z_\gal+\nu (z_\gal + y_\gal)
    \int_0^1dtt e^{it(zy)} k \right) \,,
\ee
\be
\label{tilw}
   W_0 (z,y;\psi_{1,2},k | x) = W_0 (\tilde{y};\psi_{1,2},k | x) \,,
\ee
where
\be
\label{tilde y}
   \tilde y_\gal =y_\gal+\nu (z_\gal + y_\gal)\int_0^1dt(t-1)
   e^{it(zy)}k
\ee
are the elements with the defining property
$[\tilde y_\gal\,,\,S_{0\beta}]_*=0$.
An arbitrary ``function" $W_0 (\tilde{y};\psi_{1,2},k | x)$
on the r.h.s. of (\ref{tilw}) contains star-products of
$\tilde{y}_\gal$. It is important that $\tilde{y}_\gal$
obey the deformed oscillator algebra commutation
relations (\ref{modosc}),
\be
\label{y com} [\tilde y_\gal,\tilde y_\beta]_*=
       2i\epsilon_{\gal\beta}(1+\nu k)\,,\quad \tilde y_\gal k=-k
       \tilde y_\gal \,.
       \ee

Eqs. (\ref{B_0})-(\ref{tilw}) solve all the equations
(\ref{WW}), (\ref{SS}) except for
\be
\label{vaw}
    dW_0=W_0 *\wedge W_0  \,,
\ee
which requires a further specification of $W_0$.
An appropriate ansatz is
\be
\label{W_0}
    W_0=\omega_0+\lambda h_0\psi_1\,,\quad
       \omega_0=\frac1{8i}\,\omega_0^{\gal\beta} (x)
                \{\tilde y_\gal, \tilde y_\beta\}_* \,,\quad
       h_0=\frac1{8i}\, h_0^{\gal\beta}(x)
                 \{\tilde y_\gal, \tilde y_\beta\}_*\,,
\ee
where $\omega_0^{\gal\beta} (x)$ and $h_0^{\gal\beta}(x)$
are identified with Lorentz connection and dreibein of the background
space and are required to solve (\ref{vaw}).
Here  the properties of the deformed oscillators
(\ref{y com}) play a crucial role, guaranteeing that the
anticommutators $\{y_\gal ,y_\beta \}_*$=
$y_\gal *y_\beta$+$y_\beta *y_\gal$ satisfy the
$sp(2)$ commutation relations for all $\nu$ \cite{Op}.
As a result, the gauge fields (\ref{W_0}) take values
in the $d3$ AdS algebra $o(2,2)\sim sp(2)\oplus sp(2)$ and
 (\ref{vaw}) describes AdS background.

Once a vacuum solution is known, one can study the system
(\ref{WW}), (\ref{SS})
perturbatively expanding the fields as
\be
\label{pert}
   B=B_0+B_1+\ldots \,,\qquad
   S_\gal=S_{0\gal}+S_{1\gal}+\ldots \,,\qquad
   W=W_0+W_1+\ldots \,.
\ee
Substitution of these expansions
into (\ref{WW}), (\ref{SS}) gives in the lowest order
\be
\label{WW_1}
   D_0\,W_1=0\,,
\ee
\be
\label{WB_1}
   D_0\,C=0\,,
\ee
\be
\label{WS_1}
   D_0\, S_{1\gal}=[W_1\,,\,S_{0\gal}]_* \,,
\ee
\be
\label{SS_1}
   [S_{0\gal}\,,\,S_{1\beta}]_*-[S_{0\beta}\,,\,S_{1\gal}]_*=
      -2i\epsilon_{\gal\beta}\,C*K\,,
\ee
\be
\label{SB_1}
   [S_{0\gal}\,,\,C]_*=0 \,,
\ee
where we denote $C=B_1$ and $D_0$ is the background covariant
derivative that acts on a $r$-form $P$ as
$ D_0\,P=dP-W_0\wedge P+(-)^r P\wedge W_0 $.

To analyze the system (\ref{WW_1})-(\ref{SB_1}) one proceeds
as follows. {}From (\ref{SB_1}), one concludes that
$C$ has a form similar to (\ref{tilw}),
i.e. $C=C (\tilde{y};\psi_{1,2},k | x)$.
Expanding $C$ as
$   C=C^{aux}(\tilde{y};\psi_1,k | x)
    +C^{dyn}(\tilde{y};\psi_1 ,k | x)\psi_2 $,
one identifies \cite{BPV} $C^{aux}$ with some topological fields
which carry no degrees of freedom, and $C^{dyn}$ with the
generating function for the spin 0 and spin 1/2 matter fields.
Namely, in accordance with the normal spin-statistics,
$\tilde{y}$-even (odd) part of $C^{dyn}$ identifies with
the generating
function for spin 0 (1/2) matter fields, along with all their
on-mass-shell non-trivial derivatives \cite{BPV}.
The equation (\ref{WB_1}) amounts to free field equations.
Resolving the constraints (\ref{SS_1}),
one reconstructs the auxiliary field $S_{1\gal}$ as a linear
functional of $C$, $S_{1\gal}=S_{1\gal}(C)$, up to a gauge ambiguity.
Then, (\ref{WS_1}) allows one to express a part of degrees of freedom
in $W_1$ via $C$, while the rest modes, which belong to the kernel
of the mapping $[S_{0\gal}\,, \dots]_*$, remain free.
These free modes are again arbitrary functions of
$\tilde{y}_\gal$, i.e.
\be
\label{omtil}
   W_1 = \omega (\tilde{y};\psi_{1,2},k | x) + \Delta W_1 (C) \,,
\ee
where $\omega (\tilde{y};\psi_{1,2},k | x)$ corresponds to the
HS gauge fields, and the dynamical equations
for them are imposed by eq.~(\ref{WW_1}) after
(\ref{WS_1}) is solved.
Eq.~(\ref{WW_1}) describes the $C$-dependent first order corrections
to the HS strengths for $\omega$, which are argued below to vanish.
In principle, one can proceed similarly in the highest orders.
However, the computation complicates enormously in the second
order, and one needs some efficient methods to proceed.

\section{Integrating Flow}

A remarkable property of eqs. (\ref{WW}), (\ref{SS}) is that
they admit a flow which allows one
to express constructively  solutions of the full system in terms of
free fields. Since our perturbation expansion is just an expansion
in powers of the physical fields which are identified
with the deviation $C$ of $B$ from its vacuum value
$\nu$, let us introduce a formal perturbation expansion
parameter $\eta$ (i.e. the coupling constant) as follows
\be
\label{fr}
   B(\eta)=\nu+\eta {\cal B}(\eta) \,.
\ee
Simultaneously, the rest of the  fields acquire a formal
dependence on $\eta$, $W=W(\eta)$ and $S_{\gal}=S_{\gal}(\eta)$.
The system (\ref{WW}), (\ref{SS}) takes a form
\be
\label{WWe}
   dW=W*\wedge W  \,,\quad d{\cal B}=W*{\cal B}-{\cal B}*W \,,\quad
   dS_\gal=W*S_\gal-S_\gal*W  \,,
\ee
\be
\label{SSe}
   S_\gal*S^\gal=-2i(1+\nu K+\eta {\cal B}*K) \,,\quad
   S_\gal*{\cal B}={\cal B}*S_\gal \,.
\ee
Now, one observes that for the limiting case $\eta=0$ the system
(\ref{WWe}), (\ref{SSe}) reduces to the free one. Indeed, setting
$$
   \nu=B_0\,,\quad {\cal B}(0)=B_1\equiv
   C\,,\quad W(0)=W_0\equiv \omega\,, \quad
   S_\gal(0)=S_{0\gal} \,,
$$
we see that at  $\eta=0$, the system (\ref{WWe}),
(\ref{SSe}) amounts to $S_\gal =S_{0\gal}$,
$W = \go(\tilde{y};\psi_{1,2},k | x)$,
${\cal B} = C(\tilde{y};\psi_{1,2},k | x)$,
 $d\go = \go *\wedge \go$ and $dC=\go * C-C * \go$. The latter two
equations describe vacuum background gauge fields and
free matter field equations, respectively.
This situation is similar to that with contractions
of Lie algebras.
For all values of $\eta \neq 0$, the systems of equations
(\ref{WWe}), (\ref{SSe}) are pairwise equivalent since the field
redefinition (\ref{fr}) is non-degenerate. On the other hand,
although the field redefinition (\ref{fr}) degenerates at $\eta =0$,
eqs.~(\ref{WWe}), (\ref{SSe}) still make sense for $\eta=0$.
This limiting system describes the free field dynamics.

Remarkably, the two inequivalent systems are still related to
each other. To show this let us define a flow with respect to $\eta$
as follows:
\be
\label{ps 1}
   \frac{\partial W}{\partial\eta}=
     (1-\mu)\;{\cal B}*\frac{\partial W}{\partial\nu}
     +\mu\; \frac{\partial W}{\partial\nu}*{\cal B} \,,
\ee
\be
\label{ps 2}
  \frac{\partial {\cal B}}{\partial\eta}=
   (1-\mu)\;{\cal B}*\frac{\partial {\cal B}}{\partial\nu}
   +\mu\; \frac{\partial {\cal B}}{\partial\nu}*{\cal B}   \,,
\ee
\be
\label{ps 3}
   \frac{\partial S_\gal}{\partial\eta}=
    (1-\mu)\; {\cal B}*\frac{\partial S_\gal}{\partial\nu}
    +\mu\; \frac{\partial S_\gal}{\partial\nu} * {\cal B} \,,
\ee
where $\mu$ is an arbitrary parameter.
This flow is compatible with
(\ref{WWe}), (\ref{SSe}). Therefore, solving the system
(\ref{ps 1})-(\ref{ps 3}) with the initial data
${\cal B}(\eta=0)=C\,,\, W(\eta=0)=\omega\,,\,
S_\gal(\eta=0)=S_{0\gal}$,
we can express solutions of the full nonlinear system at $\eta =1$
via solutions of the free system at $\eta =0$.
The existence of the integrating flow
(\ref{ps 1})-(\ref{ps 3}) takes its origin in the fact that
from the viewpoint of the system (\ref{WW}), (\ref{SS}),
$B$ behaves like a constant: it commutes to $S_\gal$ and
satisfies covariant constancy condition. Knowledge of the
vacuum solution with $B=\nu $
can be used to reconstruct the full dependence
of $B$. Indeed, the
qualitative meaning of (\ref{ps 1})-(\ref{ps 3})
is that a derivative with respect to $\eta{\cal B}$ is the same
as that with respect to $\nu$.
(All fields acquire a non-trivial
dependence on $\nu$ via the vacuum solution as it follows e.g.
from eqs. (\ref{omtil}), (\ref{tilde y}).)
The flows (\ref{ps 1})-(\ref{ps 3})
at different $\mu$ are equivalent modulo gauge transformations
\cite{PV}.

This approach allows one to derive the relevant field redefinitions
order by order since the r.h.s.-s
of (\ref{ps 1})-(\ref{ps 3}) contain one extra power of ${\cal B}$.
In particular, one can easily derive in the first order
a field redefinition necessary to show that the
HS gauge field strengths do not admit nontrivial sources linear in
fields. In the first order
this field redefinition is local.
This result is expected since in the lowest order
the non-trivial r.h.s.-s of the equations for HS gauge
fields are  the HS currents  bilinear in the matter fields.

Remarkably, the method works in all higher orders,
thus reducing the full non-linear problem to the free one.
The point, however, is that beyond the first order
one has to be careful in making statements
on the locality of the mapping induced by the flow
(\ref{ps 1})-(\ref{ps 3}).
Actually, although it does not
contain explicitly space-time derivatives, it contains them
implicitly via highest components $C_{\gal_1 \ldots \gal_n}$
of the generating function $C(\tilde{y})$ which are identified
\cite{BPV} with the highest derivatives of the matter
fields due to the equations (\ref{WB_1}).
For example, at $\mu=0$ in the second order in $C$ one gets
$ \frac{\partial}{\partial\eta}B_2(z,y)=
C(\tilde y)*\frac{\partial}{\partial\nu}C(\tilde y) $.
Because of the properties of the
$*$-product, for each fixed rank
multispinorial component of the l.h.s. of this formula,
its r.h.s. is an infinite
series involving bilinear combinations of the components
$C_{\gal_1 \ldots \gal_n}$
with all $n$. Therefore, the r.h.s.-s of (\ref{ps 1})-(\ref{ps 3})
effectively contain all orders of the space-time derivatives,
i.e. the  mapping resulting from (\ref{ps 1})-(\ref{ps 3})
describes some non-local transformation. This means that one cannot
treat the system (\ref{WW}), (\ref{SS}) as locally equivalent
to the free system.

To illustrate this issue it is instructive to consider an example
of some matter field $C$ interacting with the gravitational field
fluctuating near the AdS vacuum solution.
Schematically, the mechanism is as follows.
Linearized Einstein equations have a form
(with appropriate gauge fixings)
\be
\label{lein}
( L^C -\Lambda^2 ) h_{\mu\nu}=T_{\mu\nu}(C)  \,, %\E
\ee
where $h_{\mu\nu}$ is the fluctuational part of the metric tensor,
$L^C$ is the linear operator corresponding to the l.h.s.
of the free field equations of the matter fields $L^C C =0$,
while $\Lambda = \gal \lambda $ with some numerical coefficient
$\gal\neq 0$.
It is important that when the cosmological constant is non-vanishing,
the term with $\Lambda^2$ turns out to be non-vanishing too.
This property allows one to solve formally (\ref{lein})
by a field redefinition
\be
\label{nonl h}
    h_{\mu\nu}'=h_{\mu\nu}-( L^C -\Lambda^2 )^{-1} T_{\mu\nu}(C)=
    h_{\mu\nu}+ \Lambda^{-2} \sum_{n=0}^\infty (\Lambda^{-2} L^C)^n
    T_{\mu\nu}(C) \,.
    \ee
Clearly, a non-vanishing dimensionful
constant, the cosmological constant, plays an important role in this
analysis. It can be shown \cite{PV} that this field redefinition
admits a natural realization in terms of the generating function $C$
in agreement with the general analysis above.

\section{$N=2$ Supersymmetry and Truncations}\label{GlobSym}

The full system (\ref{WW}), (\ref{SS}) is explicitly
invariant under the HS gauge transformations
(\ref{delta W}). Fixation of the vacuum solution (\ref{B_0}),
(\ref{S_0}), (\ref{W_0}) breaks this local symmetry down to some
global symmetry, the symmetry of the vacuum. This global
symmetry is generated
by the parameter $\gvep_{gl}(x)$ obeying the conditions
\be
\label{de}
    d\gvep_{gl}=[W_0\,,\,\gvep_{gl}]_* \,,
\ee
\be
\label{eS}
    [\gvep_{gl}\,,\,S_{0\alpha}]_* = 0 \,,
\ee
which follow from the requirement that
$\delta W_0=0$, $\delta S_{0\alpha}=0$
under the transformations (\ref{delta W})
($\delta B_0=0$ holds automatically for $B_0 =const$).
The condition (\ref{eS}) implies that
\be
   \gvep_{gl}(z,y;\psi_{1,2},k | x) =
   \gvep_{gl} (\tilde{y};\psi_{1,2},k | x).
\ee
The dependence of $\gvep_{gl}$ on the space-time coordinates $x_\mu$
is fixed by the differential equation (\ref{de})
with arbitrary initial data $\gvep_{gl}(x_0)=\gvep_{gl}^0$
at any space-time point $x_0$.
This infinite-dimensional global symmetry is also the symmetry of
the linearized system (\ref{WW_1})-(\ref{SB_1}).

The full global symmetry algebra contains elements
$\gvep_{gl}$ linear in $\psi_2$.
This part of the symmetry mixes
the matter and topological modes in $C$ and
does not allow unitary realization on quantum states. %\cite{Fut}.
We therefore analyze the subalgebra $A^g$
of the full global symmetry algebra with the
$\psi_2$-independent parameters
\footnote{Also we factor out a trivial
central element  corresponding to a constant parameter
$\gvep_{gl} (\tilde{y};\psi_1 ,k | x) $=
$\gvep_{gl} $. }.

The algebra $A^g$ is infinite-dimensional due to the dependence
of $\gvep_{gl} (\tilde{y};\psi_1, k | x) $ on
$\tilde{y}_\alpha$. A maximal finite-dimensional
subalgebra of $A^g$, $osp(2,2)\oplus osp(2,2)$, is
spanned by the generators $\Pi_{\pm}T^A$, where
$\Pi_{\pm}=\frac{1}{2}(1\pm\psi_1)$ and
$T^A=\{T_{\alpha\beta} \,,\, Q_\alpha^{(1)} \,,\,
Q_\alpha^{(2)} \,,\, J\}$,
\be
  T_{\alpha\beta} = \frac1{4i}\{\tilde y_\alpha,\tilde y_\beta \}_*
  \,,\quad
  Q_\alpha^{(1)} = \tilde y_\alpha     \,,\quad
  Q_\alpha^{(2)} = \tilde y_\alpha k   \,,\quad
  J = k+\nu \,.
\ee
The fact that the generators $T^A$ close to
$osp(2,2)$ was shown in \cite{BWV}.

As shown in \cite{BPV}, the dynamical components
$C^{dyn}(\tilde y;k,\psi_1|x)\,\psi_2$
decompose into four bosonic and four fermionic
infinite-dimensional representations of the AdS algebra $o(2,2)$,
each describing a single AdS particle.
These free fields form altogether
an irreducible $d3$ $N=2$ hypermultiplet
constituted by 4 scalar and 4 spinor fields
\footnote{Note that this complexified multiplet is formally reducible
with the irreducible subsets singled out by the projectors
$\Pi_{\pm}$. However, these subsets turn out to be complex conjugated
to each other after imposing appropriate reality conditions
(see \cite{PV} for details).},
\be
\label{hypm}
   \left\{ \; C^0_+(x)\,,\,C^0_-(x)\,,\, C^1_+(x)\,,\,C^1_-(x)\,,\;
   C^0_{+\gal}(x)\,,\,C^0_{-\gal}(x)\,,\, C^1_{+\gal}(x)\,,\,
   C^1_{-\gal}(x)\; \right\}\,.
\ee
Here we use the following expansion of $C^{dyn}(\ty;k,\psi_1)$,
\be
\label{Cd ex}
   C^{dyn}(\ty;k,\psi_1)=[C^0_+(\ty)+C^0_-(\ty)]+
      [C^1_+(\ty)+C^1_-(\ty)]\psi_1 \,,
\ee
where $C_\pm=P_\pm C$, $P_\pm=\frac{1\pm k}2$.
Thus, the proposed field equations describe HS
interactions of a $N=2$ massive hypermultiplet.
The values of mass are related to the parameter $\nu$
as follows,
\be
\label{Mb}
    M^2_\pm =\gl^2\frac{\nu(\nu\mp 2)}2
\ee
for bosons, and
\be
\label{Mf}
    M^2_\pm  =\gl^2\frac{\nu^2}2
\ee
for fermions \cite{PV}. Here $\pm$ originates from
$C_\pm$. The doubling of fields of the same mass
is due to  the operator $\psi_1$.

In the massless case $\nu=0$ there exists \cite{PV}
a truncation induced by some involutive symmetry
of the system (\ref{WW}), (\ref{SS}), that preserves $N=2$ SUSY.
This symmetry is based on the automorphism $k\to -k$ and
the antiautomorphism $\gs$ \cite{Ann, KV1},
\be
\label{sigma}
    \gs [A(z,y;\psi_{1,2},k,\rho)] =
            A^{rev}(-iz,iy;\psi_{1,2},k,\rho) \,.
\ee
(Here the notation $A^{rev}(...)$ means that an order of all
product factors in the monomial expressions on the r.h.s.
of (\ref{gexp}) is reversed.) The reduced $N=2$ massless
supermultiplet consists of
two bosonic and two fermionic fields \cite{PV},
\be
\label{supm N2}
   \left\{ \; C^0_1(x)\,,\, C^1_0(x)\,,\,
   C^0_{0\gal}(x)\,,\, C^0_{1\gal}(x) \;  \right\}
\ee
(we  use the convention $C(k)= C_0 + C_1 k$).
This additional reduction compatible with $N=2$ SUSY is
a manifestation of the well-known shortening of massless
supermultiplets.

There exists \cite{PV} also an alternative truncation of the system
based on the antiautomorphism $\gs$ (\ref{sigma}), that breaks
$N=2$ SUSY down to $N=1$ SUSY $osp(1,2)\oplus osp(1,2)$
with the generators
\be
\label{N1}
   T_{\pm,\,\gal\gb} = \frac1{4i}\Pi_\pm
      \{\ty_\gal,\ty_\gb \} \,,\quad
   Q_{\pm,\,\gal} = \Pi_\pm  \ty_\gal
\ee
and makes sense for  arbitrary mass (i.e. $\nu$).
The truncated $N=1$ matter supermultiplet contains
the following 2 scalars and 2 spinors \cite{PV},
\be
\label{supm}
   \left\{ \; C^0_+(x)\,,\; C^0_-(x)\,,\;
    C^0_{1\,\gal}(x) \,,\; C^1_{0\,\gal}(x) \; \right\} \,,
\ee
with the masses (\ref{Mb}) and (\ref{Mf}).
In the massless case $\nu=0$, one can perform a further
truncation and obtain a shortened $N=1$ supermultiplet
 containing
one scalar and one spinor massless field.

\section{Inner symmetries and $N$-extended SUSY}\label{YM}

An important fact about HS dynamical systems
is that they admit a natural extension to the case with non-Abelian
internal (Yang-Mills) symmetries, as was first discovered
in the $d4$ case in \cite{Ann,KV1} and analyzed in detail
for the $d3$ case in \cite{PV}.
The key observation is that the system (\ref{WW})-(\ref{SS})
remains consistent if components of all fields take their values
in an arbitrary associative algebra $M$ with a unit element $I_M$,
i.e. the fields $W$, $B$, and $S_\gal$ take  values
in $A^{ext} = A\otimes M$, where $A$
is the associative algebra with the general element (\ref{gexp}).
The gravitational sector is associated with
$A\sim A\otimes I_M$ and commutes with
$M\sim I_A\otimes M$,
where $I_A$ is the unit element of $A$.
Therefore, $M$ describes internal symmetries in the model.
For the case of semisimple finite-dimensional inner symmetries,
$M$ is identified with some matrix algebra $M=Mat_n $,
i.e.
$$
   W(z,y;\psi_{1,2},k)\to W_i{}^j(z,y;\psi_{1,2},k) \,,\quad
   B(z,y;\psi_{1,2},k)\to B_i{}^j(z,y;\psi_{1,2},k) \,,
$$
$$
   S_\gal(z,y;\psi_{1,2},k,\rho)\to
   S_{\gal,\,i}{}^j(z,y;\psi_{1,2},k,\rho)  \,.
$$

The extended systems describe \cite{PV} HS
interactions of the $N=2$ hypermultiplets in
the representations
\footnote{We use the convention that fields in $k\otimes \bar{l}$
are complex conjugated to those in $l\otimes \bar{k}$.}
$n\otimes \bar{m} \oplus m\otimes \bar{n}$
of $u(n)\oplus u(m)$ Yang-Mills symmetries.
There exist consistent truncations of these extended systems
based on the antiautomorphism $\gs$ (\ref{sigma}) and appropriate
antiautomorphisms of $Mat_n$, which break $N=2$ SUSY down
to $N=1$ SUSY. The reduced systems describe
$N=1$ matter supermultiplets in the representations $n\otimes m$
of $o(n)\oplus o(m)$ or $usp(n)\oplus usp(m)$ internal symmetries
(for more detail see \cite{PV}).

Finally let us discuss  extended supersymmetries
with $N>2$. In \cite{OP1}, it was shown that there is a simple way to
incorporate $N$-extended superalgebras $osp(N,2m)$
($m=1$ for the $d=3$ case under consideration) by supplementing
the bosonic generating elements of the Heisenberg algebra
$y_\alpha$ with the Clifford elements $\phi^i$ ($i=1,\ldots, N$).
In this approach the Clifford algebra $C_N$ is a particular case of
the matrix algebra $Mat_{2^{\frac{N}{2}}}$ (for $N$ even),
while the generators of the $osp(N,2m)$ are
realized in terms of the bilinears
\be
\label{genext}
   T_{\alpha\beta} = \{y_\alpha , y_\beta \} \,, \qquad
   Q^i_\alpha = y_\alpha \phi^i \,,\qquad M^{ij} = [\phi^i ,\phi^j ]
   \,,
\ee
provided that
\be
   [y_\alpha, y_\beta ] = 2i \epsilon_{\alpha\beta} \,,\qquad
   \{\phi^i, \phi^j \} = 2\delta^{ij} \,.
\ee

Now we observe that this construction is not working for the deformed
oscillators because the generators (\ref{genext}) do not form
a closed algebra if $y_\alpha$ is replaced by $\hat{y}_\alpha$
(\ref{modosc}) with $\nu\neq 0$.
This result can be explained as follows. For $\nu \neq 0$ the mass of
the matter supermultiplet is non-vanishing and the spin range within
a supermultiplet increases with $N$. Since
massive fields of spins greater than 1/2 are not included
in our model,
$N>2$ extended supersymmetry cannot be realized.
In the massless case, however, one can realize higher supersymmetries
within only scalar and spinor fields \cite{dwtol} due to
trivialization of
the notion of spin for the $d=3$ massless case (i.e. trivialization
of the $d=3$ massless little group).

Thus we conclude that the model under consideration admits $N > 2$
extended supersymmetry only for the massless vacuum $\nu = 0$.

\section{Conclusions}

The main conclusion is that dynamical systems
based on infinite sets of HS gauge fields admit interesting
structures which allow their constructive perturbative solvability.
The fact that some system can be integrated order by order with the
help of a non-local field redefinition is  not surprising.
What is special about the HS systems is that such a field
redefinition
is described in a systematic way by some flow with respect to an
additional evolution parameter. As a result, one can reconstruct
solutions of the non-linear HS equations in terms of those of
the free system  by integrating ordinary differential equations.
We believe that this fact can be interpreted as some sort of
integrability
of the $d=3$ HS equations, although a rigorous proof of this
statement remains to be elaborated. Moreover, the very concept of
integrability may need to be modified in application to HS models.
Indeed, as mentioned at the end of sect. 2, the part of the equations
that contain space-time derivatives has a form of zero-curvature
conditions and can be integrated explicitly. As a result, one is
to solve only the constraint
part of the system (in some analogy with Hamiltonian reduction).
The latter problem is not of the evolution type
however, because the equations (\ref{SS}) are some
integral equations in the auxiliary spinor space in which the star
product acts. The main result reported in this paper is how
the problem of solving these constraints
is reduced to solving ordinary
differential equations  with respect to
the coupling constant $\eta$ provided that a
particular solution with $B=\nu=const$ is found.

We argue that the resulting field redefinition
 is essentially non-local in the space-time sense.
A non-local character of the transformation manifests itself
in the appearance of infinite series in the inverse cosmological
constant. Taking into account that
HS gauge interactions are known \cite{fv} to require non-analyticity
in the  cosmological constant, we conjecture that HS gauge theories
are indeed non-local in a certain sense.  This conjecture agrees with
the light-cone analysis in \cite{met} and fits the ideology of the
modern string theory.

\section*{Acknowledgments}

Authors are grateful to R.~R.~Metsaev for a useful comment.
This research was supported in part by INTAS, Grant No.96-0538
and by the RFBR Grant No.96-01-01144.
S.~P. acknowledges a partial support from Landau Scholarship
Foundation, Forschungszentrum J\"ulich.


\begin{thebibliography}{77}
\bibitem{bl}
   M.~P.~Blencowe, {\it Class. Quantum Grav.}~{\bf 6} (1989) 443.
\bibitem{more}
  M.~A.~Vasiliev, {\it Phys.~Lett.} {\bf B285} (1992) 225
\bibitem{U}
   R.~Utiyama, {\it Phys. Rev.} {\bf 101} (1956) 1597.
\bibitem{Kib}
   T.~W.~B.~Kibble, {\it J. Math. Phys.} {\bf 2} (1961) 212.
\bibitem{ChW}
   A.H.~Chamseddine and P.~West, {\it Nucl.~Phys.}
   {\bf B129} (1977) 39.
\bibitem{MM}
   S.~W.~MacDowell and F.~Mansouri, {\it Phys. Rev. Lett.}
   {\bf 38} (1977) 739.
\bibitem{SW}
   K.~Stelle and P.~West, {\it Phys.~Rev.} {\bf D21} (1980)  1466.
\bibitem{W}
   E.~Witten, {\it Nucl.~Phys.} {\bf B311} (1989) 46.
\bibitem{Op}
   M.~A.~Vasiliev, {\it JETP Lett.} {\bf 50} (1989), 374;
   {\it Int.~J.~Mod.~Phys.} {\bf A6} (1991) 1115.
\bibitem{un} M.A.~Vasiliev, {\it Class.~Quant.~Grav.}
   {\bf 11} (1994) 649.
\bibitem{Eq} M.~A.~Vasiliev, {\it Mod.~Phys.~Lett.} {\bf A7}
   (1992) 3689.
\bibitem{PV}
   S.~F.~Prokushkin and M.~A.~Vasiliev, {\it Nucl.~Phys. B}
   (to be published), {\tt hep-th/9806236}.
\bibitem{BPV}
   A.~V.~Barabanschikov, S.~F.~Prokushkin, and M.~A.~Vasiliev,
   {\it Rus. Theor.~Math.~Phys.} {\bf 110} (1997)  295,
   {\tt hep-th/9609034}.
\bibitem{BWV}
   E.~Bergshoeff, B.~de~Wit, and M.~A.~Vasiliev,
   {\it Nucl.~Phys.} {\bf B366} (1991) 315

\bibitem{Ann}
   M.~A.~Vasiliev, {\it Ann. Phys.} (N.Y.) {\bf 190} (1989) 59.
\bibitem{KV1}
   S.~E.~Konstein and M.~A.~Vasiliev, {\it Nucl. Phys.\/}
   {\bf B331} (1990) 475.

\bibitem{OP1} M.~A.~Vasiliev, {\it Fortschr. Phys.} {\bf 36}
   (1988) 33.
\bibitem{dwtol} B.~de~Wit, A.~K.~Tollst\'en, and H.~Nicolai,
              {\it Nucl.~Phys.} {\bf B392} (1993) 3, hep-th/9208074.

\bibitem{fv}
    E.~S.~Fradkin and M.~A.~Vasiliev,
    {\it Phys. Lett.} {\bf B189} (1987) 89.
\bibitem{met}
    R.~R.~Metsaev, {\it Mod.~Phys.~Lett.} {\bf A4} (1991) 359.

\end{thebibliography}
\end{document}